# Introducing convex layers to the Traveling Salesman Problem.

Liew Sing
liews_ryan@yahoo.com.sg

March 19, 2012

**Abstract**

In this paper, we will propose convex layers to the Traveling Salesman Problem (TSP). Firstly, we will focus on human performance on the TSP. Experimental data shows that untrained humans appear to have the ability to perform well in the TSP. On the other hand, experimental data also supports the hypothesis of convex hull i.e. human relies on convex hull to search for the optimal tour for the TSP. Secondly, from the paper published by Bonabeau, Dorigo and Theraulaz, social insect behavior would be able to help in some of the optimizing problems, especially the TSP. Thus, we propose convex layers to the TSP based on the argument that, by the analogy to the social insect behavior, untrained humans' cognition should be able to help in the TSP. Lastly, we will use Tour Improvement algorithms on convex layers to search for an optimal tour for a 13-cities problem to demonstrate the idea.

*Keywords*: Travelling Salesman Problem, Onion-Peeling algorithms, convex layers

**1.Introduction**

Despite the fact that the Traveling Salesman Problem (TSP) is very intuitive and easy to state, it is one of the most widely studied NP-hard combinatorial optimization problem[14]. The following are the statements of the problem. A salesman is required to visit each of *n* given cities once and only once, starting from any city and returning to the original city of departure. How should he travel in order to minimize the total travel distance?[9] The difficulty becomes obvious when one considers the number of possible tours by the method of brute force searching even for a relatively small number of cities *n*. For instance, for a problem with 20 cities (*n*=20) by brute force searching, it would be (20-1)!/2 tours, which is more than $10^{18}$ tours! The TSP is a class of difficult problems whose time complexity is widely believed exponential. Any attempt to construct an algorithm for finding optimal solutions for the TSP in polynomial time (in contrast with exponential time) is also widely believed not possible. Up to date, there are two classes of algorithms in solving the TSP: *Exact* algorithms and *Approximate* (or *heuristic*) algorithms[9]. The main characteristics of Exact algorithms are guaranteed to find the optimal solution in a bounded number of steps but unfortunately also complex with codes and very demanding of computer power[9]. Examples of the most effective Exact algorithms are Cutting-Plane and Facet-Finding algorithms[9]. On the other hand, in contrast, the main characteristics of Approximate algorithms are no guarantee that optimal solutions will be found but nevertheless able to provide relatively good solutions (differs only by a few percent from the optimal solution) and these algorithms are usually have shorter running times and very simple[9]. There are three classes of Approximate algorithms[9]: 1) Construction algorithms e.g. Nearest Neighbor algorithms, which gradually construct a tour by adding a new city at each step. 2) Improvement algorithms e.g. *2-opt* algorithms, which basically based on exchanging edges[4]. And 3) Composite algorithms, which combine Construction and Improvement algorithms[9][16]. Figure 1 illustrate *2-opt* edges exchanging[4]. In addition, a generalization of *2-opt* edges exchanging to *n-opt* edges exchanging forms the basis working principles for one of the most effective Improvement algorithms in solving TSP: the Lin-Kernighan algorithm[4][11].

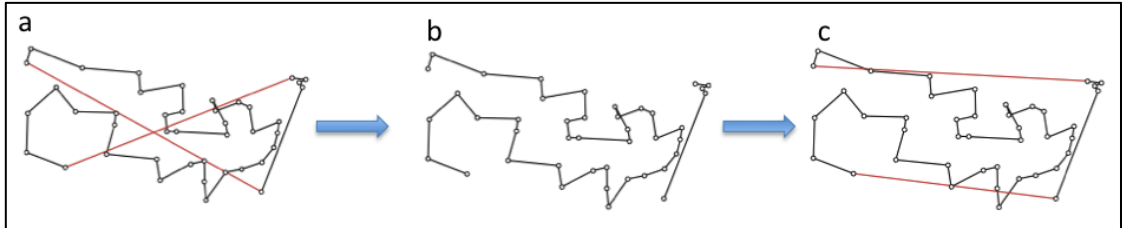

Figure 1. *2-opt* edges exchanging.

In this paper, we will propose a new approach, which is neither Exact nor Approximate: Convex layers with Improvement algorithms. In fact, this approach should be "*semi-Exact*" and "*semi-Approximate*" because, for a set of *fixed* points, there is only one type of convex layers (so called "*Exact*") and we need *Approximate* algorithms to merge the convex layers in order to obtain a tour.

The sequences of this paper are as follows. First, we will focus on the experiment results of untrained humans performance on the TSP[12]. We realized that the main reason that untrained humans appear to have the ability to perform well in the TSP is because they relied on convex hull to search for the optimal tour[13]. Second, we will focus on the Ant-Colony Optimization (ACO). Since ACO works relatively well in some of the problems, e.g. the TSP, Network routing, graph coloring, etc[1], it suggests that human cognitions may able to help in the TSP as well. Third, we focus on the modifications of convex hull to convex layers. Although convex hull serves as a good guide in a tour finding, we realized that convex layers perform even better than convex hull. In some cases, convex layers with Improvement algorithms may give us a very-close-to-optimal tour. Fourth, we will demonstrate convex layers with Improvement algorithms on a 13-cities-problem to show its feasibility as a new technique in solving the TSP. Lastly, we end the paper with conclusions.

**2. Human Performance on the TSP**
In 1996, psychological researchers MacGregor and Ormerod initiated investigations on humans performance on 10- and 20-cities TSPs[12]. In their investigations, they found that untrained humans (participants that had no particular knowledge of the TSP) were able to find solutions very close to optimal, and often draw shorter tours than simple construction algorithms[13]. In addition, MacGregor and Ormerod also concluded that participants also typically connected boundary points in order, which led MacGregor and Ormerod to propose the hypothesis of convex hull[13], and avoided crossed edges. Both features of connecting boundary points[12] and avoided crossed edges[6] are essentially for an optimal solutions to TSP. Now, we can consider convex hull as the smallest convex polygon that encloses all the points in the set i.e. the boundary points[18] or we can imagine it as how a rubber band will enclose the set of points[4]. The curve traced by the rubber band is considered as the border of the convex hull of the points. Figure 2 illustrates a convex hull, which enclosed a set of square points. Figure 3 illustrates convex layers of a set that contained nine points. The algorithm of constructing convex layers is called Onion-Peeling[10] and it is stated as follows[5]. Consider a set *S* of *n* on a 2-D plane. Compute the convex hull of *S*, and let *S'* be the set of points remaining in the interior of the hull. Then compute the convex hull of *S'* and recursively repeat this process until no more points remain.

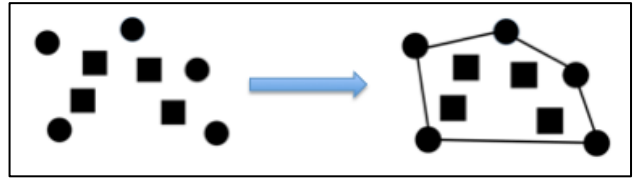
Figure 2: A convex hull with four interior points (square).

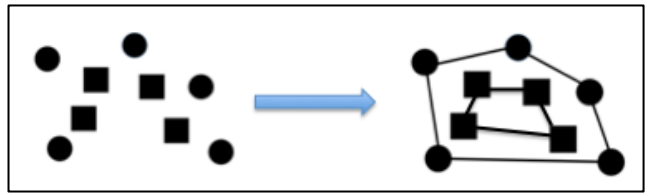
Figure 3: Convex layers of nine points.

**3. Ant-Colony Optimizations (ACO)**
In 1992, Dorigo from Belgium was inspired by the social insect behavior that the entire group of ants would communicate via pheromone trails and finds an efficient route[1]. Ever since, he started to develop the ideas based on the social insect behavior and successfully implementing the ideas into algorithms: Ant-Colony Optimizations (ACO). ACO is widely considered as one of the *Approximate* algorithms and it works as follows[4]. The algorithms work with a small group of ants moving along the edges of a given TSP problem. Each ant traces out a tour, selecting at each new vertex (point) an edge chosen among those leading to vertices not yet visited. Now, based on the selection rule, the algorithms will make use of a pheromone value associated with each edge. That is, if an edge has a high pheromone value, then it has a high probability of being selected. The algorithms will add values proportional to the lengths of the computed tours to adjust the pheromone values after the ants have all completed tours i.e. edges in good tours get their values increased more than those in poor tours. Despite the fact that ACO is very useful and has been applied effectively to many problems e.g. Network routing, graph coloring, machine scheduling, etc[1], ACO has not proved to be competitive with Lin-Kernighan-based methods in the TSP, especially on large *n*-cities problem[4].

After discussing the human performances on the TSPs that are closely related to the human basic cognitions as well as the ACO that is closely related to the social insect behavior, we thus propose a new approach in solving the TSP: convex layers with Tour Improvement algorithm.

**4. Onion-Peeling**
In 1983, Chazelle from Brow University started to study convex layers of point set *S* and he has successfully constructed an algorithm for convex layers, called Onion-Peeling, which its run time is $O(n\log n)$ for a set *S* of *n* points in $O(n)$ space[5]. As a matter of fact, convex layers were studied before Chazelle but the results were $O(n^2)$[8][17] and $O(n\log^2 n)$[15]. Note that $O$ is the so-called "big O" notation. $O(n^2)$, $O(n\log^2 n)$ and $O(n\log n)$ are called quadratic time, "n log two n" time and linearithmic time, respectively. On the other hand, $O(n^2) > O(n\log^2 n) > O(n\log n)$. In general, linearithmic time is considered "efficient" or "fast"[3].

The outlines of Onion-Peeling are as follows[5][10]. Every convex layer of set $S$, $C(S)$, can be decomposed into two convex polygonal chains, namely upper hulls and lower hulls, as illustrate in Figure 4. Let *min* and *max* denote the points with the minimum and maximum x-coordinate respectively in a convex layer. The upper hull of this layer runs clockwise and the lower hull of this layer would runs counterclockwise from *min* to *max*. Note that if the convex layer has only one or two points, the upper hull and the lower hull are the same. Assume that the set $S$ of points $p_0, p_1, \ldots, p_{n-1}$ are in non-descending order of their x-coordinates. Consider a complete binary tree $T(S)$ with leaves $p_0, p_1, \ldots, p_{n-1}$ from left to right. Let $S(v)$ denote the set of points stored at the leaves of the sub-tree rooted at node $v$ of $T$ and let $U(v)$ denote its upper hull of the convex hull of $S(v)$. Let $U(r)$ denote the upper hull in the outermost layer, where $r$ is the root of $T$. The union of all the upper hulls $U(v)$ for all nodes $v$ is a tree and called hull graph. Likewise, we construct a hull graph for the lower hull. In order to minimize the amount of space, we store the bridge, i.e. the common tangent, at each internal node $v$ connecting a point in $U(v_{min})$ and a point in $U(v_{max})$. Figure 5 illustrates the binary tree and the hull graph of $S$. The computation of the hull graph proceeds from bottom to up and computing the bridge at every node takes time linear in the number of vertices on the respective upper hulls in the left and right sub-trees. Therefore, the complexity time for the hull graph is $O(n\log n)$. The bridges computed at every node $v$ that is incident upon a vertex $p_k$ are separated into two subsets divided by the vertical line $L(p_k)$ passing through $p_k$. Those on the left are arranged in a list $\mathcal{L}(p_k)$ in counterclockwise order from the positive y direction of line $L(p_k)$. Likewise, those on the right are arranged in a list $\mathcal{R}(p_k)$ in clockwise order. Suppose the bridge at node $v$ connects vertex $p_j$ in the left sub-tree and vertex $p_k$ in the right sub-tree. The edge of $p_j$ and $p_k$ will be located at the first position in the list $\mathcal{R}(p_j)$ and $\mathcal{L}(p_k)$ i.e. edge of $p_j$ and $p_k$ is the top edge in both lists $\mathcal{R}(p_j)$ and $\mathcal{L}(p_k)$. Note that the vertices on the upper hull of the outermost layer from the hull graph beginning at the root node of $T$ can easily be retrieved. In order to compute the second layer of upper hull, we need to remove those vertices on the first layer of upper hull as well as lower hull. Removal of vertices on both upper hull and lower hull will be included in the update of the hull graph. Note that the removal of vertices can be performed in an arbitrary order. If removal of vertices on the lower hull of the hull graph are done in, say clockwise order, then the up date of the adjacency list of each vertex $p_k$ can be made easy, e.g., $\mathcal{R}(p_k)=\emptyset$. The removal of a vertex $p_k$ on the upper hull entails removal of edges incident on $p_k$ in the hull graph. Suppose $v_1, v_2, \ldots, v_{min}$ be the list of internal nodes on the leaf-to-root path from $p_k$. The edges in $\mathcal{R}(p_k)$ and $\mathcal{L}(p_k)$ removal from bottom up in $O(1)$ time each, i.e., the top edge in every list gets removed last. Figure 6 illustrates the updates of bridges when $p_{29}$ is removed.

Readers who are interested in convex hull algorithms can seek, to name a few, [2][7][8] for details.

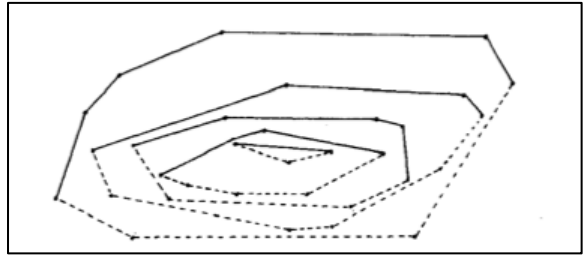

Figure 4: Upper and lower chains.

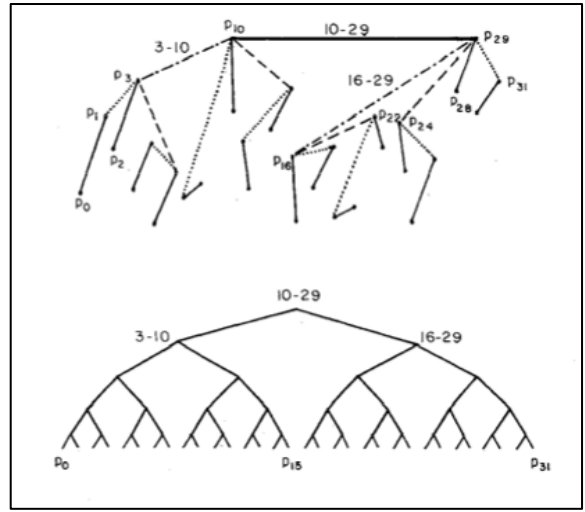

Figure 5: Hull graph of *S*.

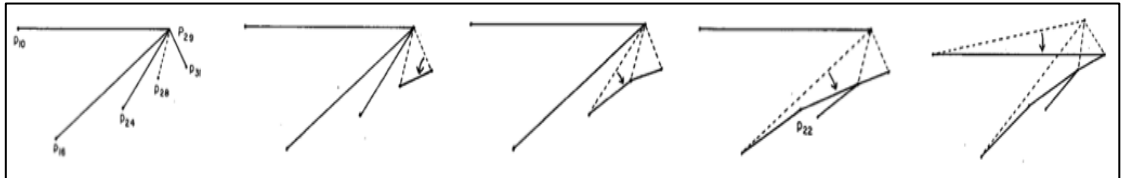

Figure 6: The removal of $p_{29}$.

**5.Convex layers with Tour Improvement algorithms.**
After introducing convex layers, we proceed to demonstrate the idea of convex layers with Tour Improvement algorithms. Tour Improvement algorithms will not be discussed as they are well established since long ago. State-of-the-art Tour Improvement algorithms are lead by Helsgaun's LKH algorithm, which its basic working principles are edge exchanging[4], and Nagata's EAX algorithm, which its basic working principles are genetic algorithms[4]. Interested readers may seek their original paper [9][14], respectively, for details. Note that in our approach, we adapt edge exchanging in our Tour Improvement algorithm because it is intuitive and it is easier to merge the convex layers with edge exchanging. On the other hand, convex layers with EAX may seem to be inappropriate because it may not able to produce a child tour by merging a convex layer with a parent tour.

We take the classic 48-States-Problem[4] as our model to demonstrate the idea. However, without the loss of generality, we only focus on the "innermost" thirteen cities (Omaha, Topeka, Kansas City, Indianapolis, Louisville, Charleston, Baltimore, Wilmington Cleveland, Detroit, Milwaukee, Chicago and Des Moines) that made up of two "innermost" convex layers. Figure 7 illustrates the convex layers for the 48-States-Problem. The outlines of the convex layers with Tour Improvement algorithms in the 48-States-Problem are as follows. First, we construct the convex layers for the problem by Onion-Peeling e.g. Figure 7. Second, we merge the convex layers (either start from the outermost or innermost layer) by Tour Improvement algorithms. In this demonstration, we use *3-opt* edge exchanging as our Tour Improvement algorithm. Lastly, iterate the Tour Improvement algorithm to merge with another convex layer until the optimal tour to be found. Figure 8 illustrates the outlines of the demonstration for the two innermost convex layers. Notice that, if we take only the thirteen cities into consideration, we obtained a very good tour.

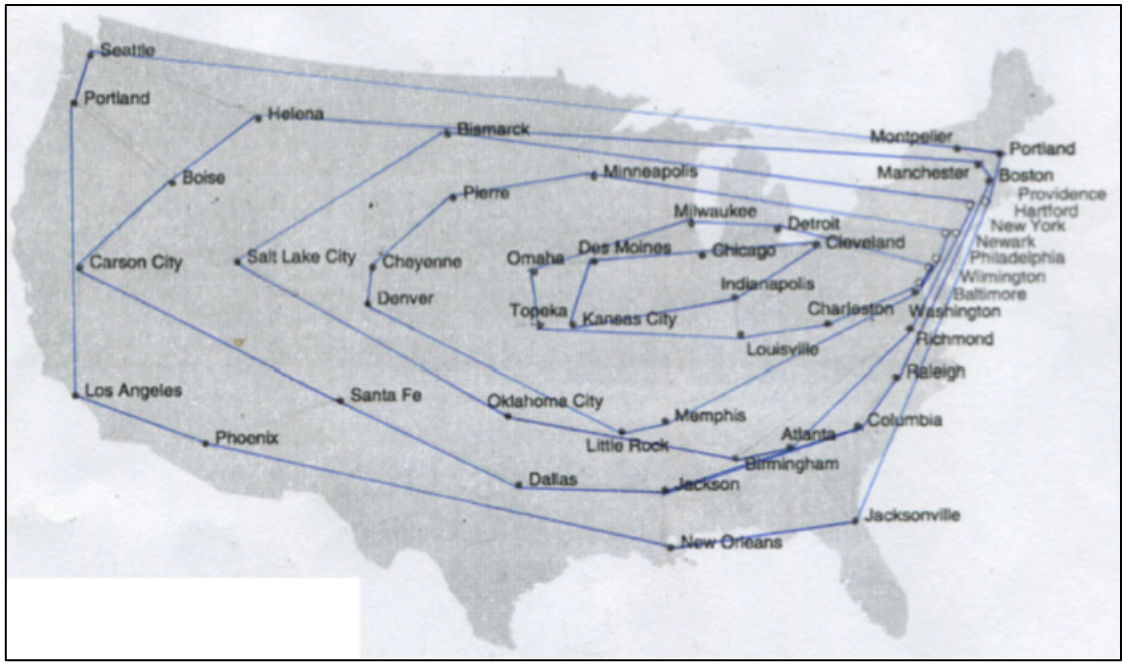

Figure 7: Convex layers of the 48-States Problem.

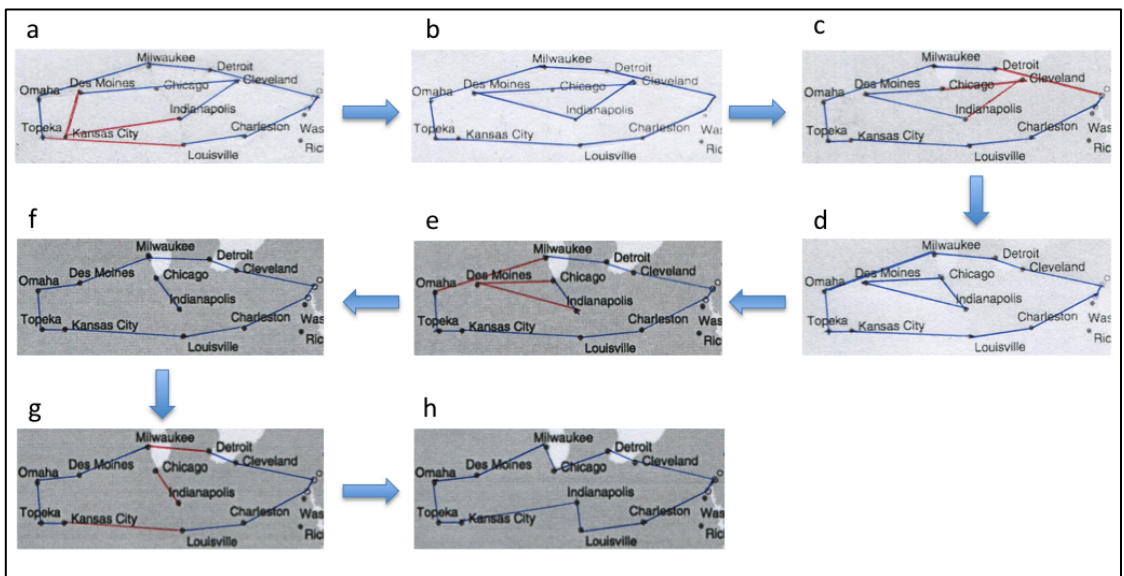

Figure 8: Convex layers with Tour Improvement algorithms (*3-opt* edge exchanging)

## 6. Conclusions

In this paper, we propose a new approach in solving the TSP, namely convex layers with Tour Improvement algorithm. The proposal was inspired by two different studies: Human performance in the TSP and Ant-Colony Optimizations (ACO). Experiment results shown that untrained humans are able to perform very well in searching an optimal tour in the TSP, and the reason behind is the cognition of convex hull (convex hull hypothesis in psychology research). Thus, in analogy to the Ant-Colony Optimizations, which were inspired by the social insect behavior, we propose convex layers with Tour Improvement algorithm.

The algorithm of convex layers (so-called onion peeling in general) is efficient and fast, i.e. $T(n)=O(n\log n)$. Together with the state-of-the-art Tour Improvement algorithm e.g. LKH, which its working principles are based on edge exchanging and is widely considered as one of the fastest Tour Improvement algorithm, we foreseen that our new approach will help in solving the TSP.